\documentclass[a4paper,11pt]{article}

\usepackage[margin=2.5cm]{geometry}

\usepackage[T1]{fontenc}
\usepackage[utf8]{inputenc}
\usepackage{microtype}
\usepackage{lmodern}
\usepackage{textcomp}
\usepackage{scalerel}
\usepackage[low-sup]{subdepth}
\usepackage[normalem]{ulem}
\usepackage{exscale}
\usepackage{verbatim}

\usepackage{amsmath}
\usepackage{amssymb}
\usepackage{mathtools}
\usepackage{dsfont}
\usepackage{physics}
\usepackage{slashed}
\usepackage{units}

\usepackage{bm} 				
\let\mathbf\bm

\usepackage[british]{babel}

\usepackage[style=english]{csquotes} 	

\usepackage{alphabeta}	
\usepackage{upgreek}

\usepackage[all,british]{foreign}	

\usepackage[affil-it]{authblk}

\usepackage{tikz}
\usepackage{tikzscale}
\usepackage{tikz-feynman}

\usepackage{cite}

\usepackage{mathrsfs}

\usepackage{relsize,exscale}
\usepackage{tensor}

\usepackage[dvipsnames]{xcolor}

\usepackage[linktocpage]{hyperref}

\usepackage[labelfont=bf]{caption}

\newcommand{\scr}{\scriptscriptstyle}

\makeatletter
\newcommand{\dalembertian}{\mathop{\mathpalette\dalembertian@\relax}}
\newcommand{\dalembertian@}[2]{%
  \begingroup
  \sbox\z@{$\m@th#1\square$}%
  \dimen0=\fontdimen8
    \ifx#1\displaystyle\textfont\else
    \ifx#1\textstyle\textfont\else
    \ifx#1\scriptstyle\scriptfont\else
    \scriptscriptfont\fi\fi\fi3
  \makebox[\wd\z@]{%
    \hbox to \ht\z@{%
      \vrule width \dimen0
      \kern-\dimen0
      \vbox to \ht\z@{
        \hrule height \dimen0 width \ht\z@
        \vss
        \hrule height 2\dimen0
      }%
      \kern-2.5\dimen0
      \vrule width 2.5\dimen0
    }%
  }%
  \endgroup
}
\makeatother

\newenvironment{acknowledgements}{\begin{abstract}}{\end{abstract}}

\newcommand{\pkcom}[1]{\textcolor{orange}{[PK:  #1]}}

\newcommand{\ils}[1]{\textcolor{blue}{#1}}
\newcommand{\ilsr}[1]{\textcolor{blue}{[IS: #1]}}

\usepackage{cancel}

\hypersetup{
    colorlinks,
    linkcolor={purple},
    citecolor={purple},
    urlcolor={purple}
}

\usepackage{orcidlink}

\numberwithin{equation}{section}

\allowdisplaybreaks						

\title{\Large\textbf{On the importance of radiation-era initial conditions\\
for tensor perturbations}}

\author[$a$]{Dra\v{z}en Glavan\,\orcidlink{0000-0002-1983-0448}\,\footnote{\href{mailto:glavan@fzu.cz}{\texttt{glavan@fzu.cz}}}}
\author[$b$]{Juraj Klari\'{c}\,\orcidlink{0000-0001-9468-5351}\,\footnote{\href{mailto:jklaric@phy.hr}{\texttt{jklaric@phy.hr}}}}
\author[$c$]{Philipp Klose\,\orcidlink{0000-0003-3702-4738}\,\footnote{\href{mailto:pklose@nikhef.nl}{\texttt{pklose@nikhef.nl}}}}
\author[$a$]{Ignacy Sawicki\,\orcidlink{0000-0003-2476-9730}\,\footnote{\href{mailto:sawicki@fzu.cz}{\texttt{sawicki@fzu.cz}}}}

\affil[$a$]{CEICO, Institute of Physics of the Czech Academy of Sciences (FZU), \newline
Na Slovance 1999/2, 182 00 Prague 8, Czech Republic}
\affil[$b$]{Department of Physics, Faculty of Science, University of Zagreb, \newline
Bijeni\v{c}ka cesta 32, 10000 Zagreb, Croatia}
\affil[$c$]{Theory Group, Nikhef, Science Park 105, 1098 XG Amsterdam, The Netherlands}

\date{}

\begin{document}

\maketitle

\begin{abstract}
Conservation of super-horizon tensor fluctuations is crucial for connecting inflation to observations.
Starting from first principles, recent works have found violations of this conservation if free-streaming radiation is produced during reheating.
We show that the non-conservation is sensitive to the radiation initial state,
and argue that the physical state should be affected by tensor perturbations that are already present during reheating.
The deviation from super-horizon conservation is then negligible, recovering the standard result from kinetic theory.
In contrast, a globally homogeneous and isotropic plasma state leads to a large suppression of tensor amplitudes.
This difference
between the local (physical) and global thermal equilibrium
settles the discrepancy between the older and recent literature.
\end{abstract}



\tableofcontents


\

\

\section{Introduction}


It is difficult to overstate the importance of primordial gravitational wave 
(GW) backgrounds as a probes of the early Universe. In particular, they 
constrain cosmic inflation, which produces
an ensemble of long wavelength scalar~\cite{Mukhanov:1981xt} and tensor~\cite{Starobinsky:1979ty}
cosmological perturbations
via gravitational particle production~\cite{Parker:1968mv}.
While scalar perturbations consistent with inflation have been observed
in the cosmic microwave background (CMB)~\cite{Planck:2018jri},
the tensor modes, i.e.~primordial gravitational waves, have so far only been constrained.
A detection would be especially significant as they
directly probe the energy scale of inflation.
Accordingly, major experimental efforts are currently underway
to reach the needed sensitivity~\cite{Hui:2018cvg, SimonsObservatory:2025avm, Belkner:2023duz, LiteBIRD:2022cnt}.

Our ability to connect inflation to tensor perturbations at the CMB relies crucially 
on understanding how they evolve on super-horizon scales.
For observationally relevant 
scales,
this includes evolution during the radiation era,
when the Universe was filled by a hot plasma.
The established literature 
holds that tensor perturbations
are frozen on 
super-horizon scales, no matter how the plasma evolves~\cite{Weinberg:2008zzc}.
In particular, this includes the hydrodynamic limit~\cite{Mukhanov:1990me}, where plasma 
efficiently equilibrates locally,
but also the free-streaming limit, where kinetic theory 
 predicts conservation~\cite{Weinberg:2003ur}.
In general, it is difficult to 
find a process 
that would significantly affect the GW background once it is produced, 
with the notable exception of damping by
free-streaming particles, such as neutrinos,
at horizon crossing \cite{Weinberg:2003ur,Baym:2017xvh}. 
However, several recent works~\cite{Ota:2023iyh,Ota:2024idm,Frob:2025sfq,Ota:2025yeu} 
have reconsidered this question in a microscopic field-theory description,
and have instead reported
that a free-streaming plasma can also
induce a non-trivial evolution of tensor perturbations
on super-horizon scales.
\\

In this work, we show that this reported super-horizon
evolution of tensor modes during the radiation era depends
sensitively on the assumed initial state of the primordial plasma.
If it equilibrates efficiently, local thermal equilibrium erases
any memory of the initial state and conservation occurs in the standard manner.
In contrast, the post-reheating initial state is
important for the results reported in
Refs.~\cite{Ota:2023iyh,Ota:2024idm,Frob:2025sfq,Ota:2025yeu,
Ota:2025rll,Ota:2026yzp}
because they
considered a plasma of free-streaming particles,
where the absence of self-interactions
should preserve information about the initial state.
We argue that the initial conditions adopted in those works are not physically well motivated, because they neglect that reheating occurs
in a universe that is already perturbed by long-wavelength tensor modes 
generated during inflation. Accounting for tensor-induced corrections 
to the initial state of the plasma drastically changes the subsequent dynamics, and the deviation from super-horizon conservation is then negligible.
Our thermal field theory results for the free-streaming case are therefore fully 
consistent with earlier kinetic-theory 
treatments~\cite{Rebhan:1994zw,Weinberg:2003ur}.

While the recent works~\cite{Ota:2025rll,Ota:2026yzp} argue
in favour of the kinetic-theory results~\cite{Rebhan:1994zw,Weinberg:2003ur},
they assert that agreement is obtained only if the initial plasma state is specified solely by global properties
of the homogeneous and isotropic background and
is taken to contain no tensor-induced perturbations inherited from inflation.
The analysis we present here reaches the opposite conclusion. 
We show explicitly that (i) including the tensor-sourced corrections to the initial 
state results in a negligible departure from super-horizon conservation of tensor modes, 
and (ii) not including the tensor-sourced corrections to the initial state yields a
huge suppression of the super-horizon tensor modes, contrary to the claims 
in~\cite{Ota:2025rll,Ota:2026yzp}. 
We also clarify why the large super-horizon growth reported in earlier 
works~\cite{Ota:2023iyh,Frob:2025sfq} does not arise in a consistent treatment, for 
different reasons in each case.
\\

The remainder of this work is structured as follows.
In Sec.~\ref{sec:framework} we outline our general strategy for computing the 
anisotropic stress, which determines in-medium corrections to the 
tensor-mode equation of motion, and discuss the role of initial conditions 
in the free-streaming limit.
In Sec.~\ref{sec: Two free-streaming plasma models}, we apply 
this framework to two specific
free-streaming plasma models, 
conformally coupled scalars and photons, and show how the predicted 
tensor mode signal crucially depends on the plasma state generated during 
reheating.
Finally, we summarize our results in Sec.~\ref{sec:conclusion},
and show that a physically motivated
initial state that includes corrections due to the
primordial tensor modes predicts conservation
of tensor perturbations on super-horizon scales. 

\section{Stochastic description}
\label{sec:framework}

In standard cosmological scenarios,
a period of reheating (see e.g. \cite{Lozanov:2019jxc} for a pedagogical review) following inflation creates the
thermal plasma that sources the expansion of the universe in the radiation era.
As is the case with any thermal system, the state of this plasma is characterized as a statistical ensemble of various possible microstates.
One convenient way to describe the dynamics of this ensemble
is to promote its microscopic degrees of freedom to stochastic variables.
The Einstein equation then relates these stochastic variables
to the geometry of space-time, which therefore also acquires a stochastic character.
On the other hand, macroscopic observables such as the scale factor of the universe, the plasma temperature, or power spectra are related to ensemble averages,
and are therefore not described by stochastic variables.

In this section, we review how to capture the impact of
the early universe plasma using this stochastic approach.
We first present the relevant equations of motion,
and then discuss the choice of initial conditions as well as their relevance in different regime,
arguing that the physical initial state after reheating should correspond to local not global equilibrium.

\subsection{Equations of motion}

We consider a universe that contains some stochastic matter field(s),
which we collectively label by~$\Phi$.
The corresponding stochastic metric tensor $g_{\mu\nu}$
evolves according to the Einstein equation
\begin{equation}
G^{\mu}{}_{\nu}[g] = \frac{\kappa^2}{2} T^{\mu}{}_{\nu}[g,\Phi]
\qquad\text{with}\qquad
\kappa^2 = 16 \pi G_{\scr \rm N} = \frac{2}{M_{\rm \scr Pl}^2}
\, .
\label{EinsteinEq}
\end{equation}
Crucially, the evolution of the matter fields depends on the geometry of space-time,
which means that they also exhibit a functional dependence on the metric tensor, $\Phi\!=\!\Phi[g_{\mu\nu}]$.
We focus on systems for which it is possible to split the metric
\begin{equation}
g_{\mu\nu} = \overline{g}_{\mu\nu} + \delta g_{\mu\nu}
\label{GeneralMetricPert}
\end{equation}
into a deterministic background contribution $\overline g$
and a small stochastic fluctuation $\delta g$.
The background metric then evolves according to the averaged Einstein equation
\begin{equation}
G{}^{\mu}{}_{\nu}[\overline{g}] = \frac{\kappa^2}{2}
	\big
\langle T^{\mu}{}_{\nu} \big[ \overline{g}{},\Phi[\overline{g}] \big] \big
\rangle
	\, ,
\label{BackgroundeinsteinEq}
\end{equation}
where the expectation value $\langle \mathcal O \rangle \equiv \tr \{ \rho \, \mathcal O \}$ 
is an ensemble average taken with respect to the matter degrees of freedom.
By subtracting Eq.~(\ref{BackgroundeinsteinEq}) from Eq.~(\ref{EinsteinEq}),
one obtains the perturbed Einstein equation
\begin{equation}
G{}^{\mu}{}_{\nu}[g] - G{}^{\mu}{}_{\nu}[\overline{g}]
= \frac{\kappa^2}{2} \Bigl( 
	T^{\mu}{}_{\nu} \big[g,\Phi[g] \big] -
	\big\langle T^{\mu}{}_{\nu} \big[ \overline g,\Phi[\overline g] \big] \big\rangle
	\Bigr) \ ,
\label{PertEinsteinEq}
\end{equation}
which yields the fundamental equations of motion for $\delta g$.
The expression on the right-hand side captures the impact of the
early universe plasma on the propagation of metric perturbations.
Our primary goal in the remainder of this paper is to evaluate
this expression at linear order in~$\delta g$.

\subsubsection*{Background}

If the background cosmology is spatially flat, homogeneous and isotropic,
its geometry is described by the Friedmann-Lema\^{i}tre-Robertson-Walker (FLRW) 
metric
\begin{equation}
\label{eq:background metric}
\overline g_{\mu\nu}(\tau) = a^2(\tau) \, \eta_{\mu\nu} \ ,
\end{equation}
where $\tau$ is the conformal time coordinate.
During the radiation era,
the expansion of the universe is sourced by quasi-thermal radiation.
The expectation value of the energy-momentum tensor~\cite{Ota:2023iyh,Frob:2025sfq}
is then diagonal and traceless.
In a weakly-coupled theory, one obtains~\cite{Kolb:1990vq,Kapusta:2006pm}
\begin{equation}
\label{eq:background stress-energy tensor}
\big\langle T^{i}{}_{i} \big[ \overline{g}{},\Phi[\overline{g}] \big] \big\rangle
=
-
\big\langle T^{0}{}_{0} \big[ \overline{g}{},\Phi[\overline{g}] \big] \big\rangle
	=
	\frac{g_{\rm eff} \pi^2T^4}{180}
	\, ,
\end{equation}
where~$g_{\rm eff}$ the effective number of radiation degrees of freedom.
If the plasma consists of a single scalar field, one has~$g_{\rm eff}=1$.
For photons, one instead has~$g_{\rm eff}=2$.%
\footnote{A more realistic cosmology with multiple 
relativistic species at reheating can have $g_\text{eff}\sim 100$,
with some degrees of freedom fermionic,
but we do not expect this to alter the overall result.}
In any case, the Friedmann equations imply
that the Hubble rate is given by
\begin{align}
H = H_\text{rh} \left(\frac{a_\text{rh}}{a}\right)^2\,, && H_\text{rh}^2 &\equiv \frac{\pi^2 \kappa^2}{180} g_{\rm eff} T_\text{rh}^4 \ , &
a \, T &= a_{\rm rh} T_{\rm rh} \ ,
\end{align}
where $T_{\rm rh}$ the reheating temperature.
We also note that, using an effective kinetic theory description,
the expectation value of the stress energy tensor is related
to the phase-space distribution functions of the various plasma particles,
\begin{align}
\big\langle T^{\mu}{}_{\nu}\big[\overline g, \Phi[\overline g] \big] \big\rangle &= \eta^{\mu\lambda} \sum_i \int \frac{\text{d}^3 \vec k}{(2\pi)^3} \frac{k_\lambda k_\nu}{\omega_i} \, f_i (\vec k) \ , &
k_\mu &= (\omega_i, \vec k) \ ,
\end{align}
where the index $i$ runs over each degree of freedom in the plasma
and $f_i (\vec k)$ is the corresponding distribution function,
which implicitly depends on the background metric $\overline g$.
The energy $\omega_i (\vec k)$ is a function of $\vec k$
that is determined by the in-medium dispersion relation of each particle.
For bosons in thermal equilibrium, the distribution function
can be identified with the Bose-Einstein distribution
\begin{align}
f_B(\omega) &= \frac{1}{e^{\, \omega/T } - 1} \ \ .
\label{BEdef}
\end{align}

\subsubsection*{Tensor perturbations}

Perturbations around the homogeneous and isotropic FLRW
background~\eqref{eq:background metric} can be decomposed into scalar,
vector, and tensor modes~\cite{Mukhanov:1990me} that do 
not couple to each other at linear order,
even when accounting for the impact of the early Universe plasma.
Our goal is to describe the linearized evolution of the tensor modes.
Hence, we set the scalar and vector modes to zero from the start,
and keep only the transverse and traceless tensor modes.
The perturbation in~\eqref{GeneralMetricPert} then reads
\begin{equation}
\delta g_{00} = \delta g_{0i} = 0 \, , \qquad
\delta g_{ij} = a^2 \kappa \, h_{ij} \, , \qquad\quad
\text{with}
\qquad\quad
h_{ii} =0 \, , \quad \partial_i h_{ij} = 0 \, ,
\label{MetricPert}
\end{equation}
where the factor of~$\kappa$ has been made explicit to ensure
that the graviton field~$h_{ij}$ has canonical dimension one. 
The perturbation of the Einstein tensor~\eqref{PertEinsteinEq} then becomes
\begin{equation}
G^i{}_j[g] - G^i{}_j[\overline{g}] = \frac{\kappa}{2} \Big( h_{ij}'' + 2\mathcal{H} h_{ij}' 
    - \nabla^2 h_{ij} \Big) + \mathcal O (h^2) \, ,
\end{equation}
where here and in the following primes denote derivatives
with respect to conformal time~$\tau$,
where~$\nabla^2=\partial_i\partial_i$ is the Laplacian,
and~$\mathcal{H}=aH$ is the conformal Hubble rate.
Note that the perturbation of the Einstein tensor
produces a massless kinetic operator for the tensor perturbation.
This is a consequence of perturbing the Einstein equation
with mixed index placement~\cite{Mukhanov:1990me}.
When perturbing the Einstein equation with both indices lowered,
one obtains an additional mass-like term that cancels with
a corresponding contribution from the right-hand side 
of Eq.~(\ref{PertEinsteinEq})~\cite{Liu:2024utl,Frob:2025sfq,Ota:2026yzp}.
Using the mixed index placement, one obtains the usual equation of motion
\begin{equation}
\label{eq:graviton eom}
h_{ij}'' + 2\mathcal{H} h_{ij}' - \nabla^2 h_{ij} = \pi_{ij} \, ,
\end{equation}
where~$\pi_{ij}$ is the {\it anisotropic stress},
defined here as the transverse-traceless projection
of the right-hand side of the perturbed Einstein equation~\eqref{PertEinsteinEq}
\begin{equation}
\pi_{ij} = \frac{\kappa}{2} \Pi_{ijkl}
    \Bigl( 
	T^{k}{}_{l} \big[g,\Phi[g] \big] -
	\big\langle T^{k}{}_{l} \big[ \overline g,\Phi[\overline g] \big] \big\rangle
	\Bigr) \, ,
\end{equation}
where
\begin{equation}
\Pi_{ijkl} = \Pi_{i(k} \Pi_{l)j} - \frac{1}{2} \Pi_{ij} \Pi_{kl} \ , 
\qquad\qquad
\Pi_{ij} = \delta_{ij} - \frac{\partial_i \partial_j}{\nabla^2}
\end{equation}
is the transverse-traceless projector.
The anisotropic stress contains two qualitatively different contributions.
The first is the stochastic noise term,
which accounts for the effect of matter fluctuations on 
top of the background spacetime~$\overline{g}_{\mu\nu}$,
\begin{equation}
\zeta_{ij} \equiv
	\kappa \, \Pi_{ijkl}
	\Big(
	T^k{}_l \big[ \overline{g}{},\Phi[\overline{g}] \big]
	-
	\big\langle T^k{}_l \big[ \overline{g}{},\Phi[\overline{g}] \big] \big\rangle
	\Big)
    \, .
\end{equation}
This contribution captures gravitational wave production from thermal fluctuations in the early universe plasma~\cite{Ghiglieri:2015nfa,Ghiglieri:2020mhm}.
This process is the dominant source of stochastic gravitational wave backgrounds in 
the standard model \cite{Ghiglieri:2020mhm,Ghiglieri:2024ghm}, and can be 
also used to e.g. constrain new physics from searches for ultra high-frequency backgrounds~\cite{Ringwald:2020ist,Ringwald:2022xif,Drewes:2023oxg}.
It is heavily suppressed on super-horizon scales,
and we do not consider it here.\footnote{For a detailed computation of the 
stochastic noise kernel in the free-streaming photon model see~\cite{Frob:2025uev}.}

If the early universe plasma is in kinetic equilibrium, and if the mean free path of all particles
in the plasma is small compared to the inverse gravitational wave frequency,
hydrodynamics predicts that
\begin{align}\label{eq:hydro perturbation}
\pi_{ij} &= \zeta_{ij}
- \kappa \, \eta \, \frac{h_{ij}^\prime}{a} 
\end{align}
where $\eta$ the shear viscosity of the early universe plasma.
Together, eqs.~\eqref{eq:graviton eom} and \eqref{eq:hydro perturbation} imply that tensor perturbations either decay rapidly or remain constant.
Importantly, the initial conditions are irrelevant because the plasma thermalizes
quickly compared to frequency of the super-horizon tensor perturbations.

However, the hydrodynamic description is not applicable
for free-streaming (including non-interacting) particles
with a mean free path that is large compared to
the inverse frequency of the gravitational waves.
In this case, initial conditions become important.
To investigate the linear propagation of tensor perturbations in this regime,
it is convenient to decompose the anisotropic stress,
as is routine in the stochastic gravity formalism~\cite{Hu:2008rga,Hu:2020luk},
into an explicit contribution
\begin{equation}
\pi_{ij}^{\rm expl.} =
\kappa \, \Pi_{ijkl} \big\langle T^k{}_l \big[\overline g + \delta g,\Phi[\overline g] \big] - T^k{}_l \big[ \overline g,\Phi[\overline g] \big] \big\rangle
\, ,
\label{ExplicitPi}
\end{equation}
that captures the fluctuations of the explicit metric dependence of the stress-energy tensor,
and an implicit contribution
\begin{equation}
\pi_{ij}^{\rm impl.} =
\kappa \, \Pi_{ijkl} \big\langle T^k{}_l \big[\overline g,\Phi[\overline g + \delta g] \big] - T^k{}_l \big[ \overline g,\Phi[\overline g] \big] \big\rangle
\label{ImplicitPi}
\, ,
\end{equation}
that captures the genuine linear response of the matter fields to metric perturbations.

\subsection{Initial conditions}
\label{subsec: Initial conditions}

In standard cosmic histories, inflation produces
a background of scalar and tensor perturbations
with wavelengths that gradually exit the causal horizon.
At the end of inflation, all fluctuations with observationally
relevant wavelengths are deep in the super-horizon regime.
Reheating is then imagined as a violent non-equilibrium process that terminates inflation
and produces the matter that drives expansion during the radiation era (see ref.~\cite{Amin:2014eta} for a review).

During reheating, matter interactions have to be
strong enough to thermalize the early universe plasma.
Crucially, they are mediated by local interactions.
We argue that the notion of thermal equilibrium is therefore inherently local.
Observationally, we note that scalar perturbations in the CMB are adiabatic
to a very high level of precision,
with strong upper bounds on isocurvature modes \cite{Planck:2018jri}.
 This implies that all species in the early universe plasma thermalized locally
on the common background of long-wavelength curvature fluctuations produced by inflation.
If the early universe plasma had instead thermalized
on the globally homogeneous and isotropic background,
it would not have inherited these pre-existing fluctuations.
Strictly speaking, existing CMB observations only 
measure the properties of scalar fluctuations, but it is difficult to imagine that
the long-wavelength tensor background would behave differently
from the scalar perturbations in this regard. 

The interactions responsible for reheating determine the largest length scale that can play a role in the physics of thermalization.
To enable efficient thermalization, this length scale has to be small compared to the causal horizon, $H^{-1}$.
If the plasma admits a particle description for hard modes with momenta $k \sim \pi T$, the scale can be identified with the mean free path of the corresponding particles.
In this case, modes with wave-lengths that are large compared to the mean free path can be described as a slowly-varying background for the particle-like modes.
In particular, the dominant effect of the long-wavelength metric perturbations is just to change the local dispersion relation of the particles.
For a massless particle in the plasma, one then obtains the perturbed mass-shell 
\begin{equation}
0 = k^2 = \overline g^{\mu\nu} k_\mu k_\nu + \delta g^{\mu\nu} k_\mu k_\nu \ .
\end{equation}
If the superhorizon perturbation only has tensor contributions,
this gives the tree-level dispersion relation \cite{Weinberg:2003ur,Baym:2017xvh}
\begin{equation}
    \omega_{\rm rh}^2 = (\delta_{ij} - \kappa h_{ij}^\text{rh})k_i k_j \, ,
\end{equation}
that fixes the energy of on-shell particles.
Hence, matter interactions drive the distribution functions of particles in the plasma to the local equilibrium shape
\begin{equation}
\label{eq:leq perturbation distribution function}
f_{\rm rh} (\vec k) 
= \frac{1}{e^{\omega_{\rm rh}/T} \mp 1} = \left[ \exp( \frac1{T} \sqrt{ ( \delta_{ij} - \kappa h_{ij}^\text{rh} ) k_i k_j} ) \pm 1 \right]^{-1} \ ,
\end{equation}
where the $+1$ applies for fermions and the $-1$ for bosons.
Crucially, this distribution function depends on the metric perturbation at the time of reheating $h_{ij}^{\rm rh}$.
Each thermalized particle species thus contributes to the implicit part of the anisotropic stress \eqref{ImplicitPi}.

The fact that the primordial plasma has to be able to thermalize at the onset of the radiation era
raises the obvious question if and to what extent the initial conditions can play a role in the subsequent evolution of tensor perturbations.
Indeed, if matter interactions remain efficient, they continue to push the distribution functions towards local equilibrium,
which erases any information about the initial state and the details of reheating.
This process also efficiently eliminates most of the anisotropic stress,
leaving only the residual damping term quoted in~\eqref{eq:hydro perturbation},
which is directly proportional to the small but finite departure of the early universe plasma from local equilibrium.

In the opposite limit, where particles cease to interact efficiently and 
start to free-stream soon after reheating, information on the initial 
state of the plasma can be preserved by evolution. In this case, the 
distribution functions of particles freeze out,
and therefore continue to depend on 
the tensor amplitude at reheating~$h_{ij}^{\rm rh}$,
while the  dispersion relation evolves together with the time-dependent
tensor perturbation~$h_{ij}$.
We focus on this free-streaming regime to determine whether
a violation of super-horizon conservation for tensor modes is possible,
and how this depends on the
initial conditions produced by reheating.
In section~\ref{sec: Two free-streaming plasma models}, we explicitly compute the resulting stress for two minimal free-streaming plasma models.
As we will demonstrate, the final result depends sensitively on the initial conditions.
\\

At this point, it is important to note that the argument we have presented above is at odds with the conclusions reached in ref.~\cite{Ota:2025rll},
where it is argued that the plasma should thermalize on the globally homogeneous and isotropic background metric, and therefore not depend on $h_{ij}$ at all.
We discuss the quantitative impact of choosing either a local or global equilibrium state in Sec.~\ref{sec:conclusion}. 

\section{Two free-streaming plasma models}
\label{sec: Two free-streaming plasma models}

In this section, we explicitly compute the anisotropic stress of two minimal models for a plasma of free-streaming particles.
First, we consider a conformally coupled, free scalar field, and then we consider a minimally coupled photon.

\subsection{Conformally coupled scalars}
\label{subsec:conformal scalars}

We consider the medium response generated by a non-interacting scalar field $\phi$ that is conformally coupled%
\footnote{We use the conformal coupling to make the computation as transparent as possible. A minimally coupled scalar produces the same leading-order results.}
to gravity. Its action is
\begin{equation}
\label{eq:scalar action}
S[\phi,g_{\mu\nu}] = \int \! d^4 x \sqrt{-g} \left( - \frac12 g^{\mu\nu} \partial_\mu \phi \partial_\nu \phi - \frac{R}{12} \phi^2 \right) \, ,
\end{equation}
where \(R\) is the Ricci scalar. The scalar sources the Einstein equation through its energy-momentum tensor,
\begin{equation}
T^\mu{}_\nu =
	\nabla^\mu \phi \partial_\nu \phi
		-
		\frac{1}{2} \delta^\mu{}_\nu g^{\rho\sigma} \partial_\rho \phi \partial_\sigma \phi
		+
		\frac{1}{6} \Big( G^\mu{}_\nu - \nabla^\mu \nabla_\nu
			+ \delta^\mu{}_\nu {\dalembertian}
		\Big) \phi^2
		\, ,
\label{EMT}
\end{equation}
and therefore contributes to the anisotropic stress in the linearized tensor-mode equation of motion~\eqref{eq:graviton eom}.
The energy-momentum tensor~\eqref{EMT}, and hence in the anistropic stress, depends on the tensor-perturbation in two ways:
Firstly, through the explicit metric dependence of the tensor itself, and secondly, through the implicit dependence arising from the scalar dynamics,
i.e. from the metric dependence of the scalar equation of motion.

The explicit contribution to the anisotropic stress~\eqref{ExplicitPi} is 
obtained by perturbing the explicit metric dependence of~\eqref{EMT} to
linear order,
\begin{equation}
\pi_{ij}^{\rm expl.}
	=
	- \frac{\kappa^2}{3a^4} \big\langle (\nabla \chi)^2 \big\rangle_{(0)} \, h_{ij}
	\, ,
\label{ScalarPiExplicit}
\end{equation}
where it is convenient to work with the rescaled field $\chi = a \, \phi$.
The supscript $(0)$ indicates that the correlator is to be evaluated at zeroth order in the tensor perturbation.
Due to the conformal coupling to gravity, $\chi$ is insensitive to the cosmological expansion at this order.%
\footnote{
It is worth pointing out that the zeroth order energy-momentum tensor, expressed in terms of the rescaled field~$\chi$, 
\begin{equation*}
T^\mu{}_\nu = \frac{1}{a^4}
    \Big[
    \eta^{\mu\rho} \partial_\rho \chi \partial_\nu \chi
    - \tfrac{1}{2} \delta^\mu_{\nu} \eta^{\rho\sigma}
        \partial_\rho\chi \partial_\sigma\chi
    +
    \tfrac{1}{6} ( \delta^\mu_{\nu} \partial^2 
        - \eta^{\mu\rho} \partial_\rho \partial_\nu )
    \chi^2
    \Big]
    \, ,
\end{equation*}
does not depend on any curvature scales apart from the overall 
prefactor~$a^{-4}$, but still differs form than the energy-momentum tensor 
of the genuine flat-space scalar field.
}
In the implicit contribution to anisotropic stress~\eqref{ImplicitPi}, many 
terms drop out after applying the transverse-traceless projector.
One finds
\begin{equation}
\pi_{ij}^{\rm impl.}
	=
	\frac{\kappa^2}{a^4} \Pi_{ijkl} 
    \big\langle \partial_k \chi \partial_l \chi \big\rangle_{(1)}
	\, ,
\label{ScalarPiImplicit}
\end{equation}
where the subscript $(1)$ indicates that this contribution is to be evaluated
at linear order in tensor perturbation.
In order to do so, one has to solve for the dynamics of the rescaled 
field~$\chi$ and specify the initial state of the plasma by imposing appropriate initial conditions. 

The equation of motion for~$\chi$, perturbed to linear order 
in the tensor perturbation, reads
\begin{equation}
\big( \partial^2 - \kappa h_{ij} \partial_i \partial_j \big) \chi = 0 \, ,
\label{ChiEom}
\end{equation}
where~$\partial^2 = \eta^{\mu\nu}\partial_\mu \partial_\nu $ is the flat space
d'Alembertian.
At super-horizon scales, we may neglect the spatial dependence of the tensor perturbations,
$h_{ij} = h_{ij}(\tau)$.
It is then useful to consider the spatial Fourier transform
\begin{equation}
\chi (\tau, \vec{k} \,)
	= \int \! d^3 \vec{x} \, e^{-i \vec{x} \cdot \vec{k}} \, \chi (\tau,\vec{x}) 
	\, ,
\end{equation}
which obeys the equation of motion
\begin{equation}
\label{eq:chi eom}
\chi^{\prime\prime} + \omega^2 \, \chi = 0 \, , 
\qquad\qquad
\omega^2 = \big( \delta_{ij} - \kappa h_{ij} \big) k_i k_j \, .
\end{equation}
At super-horizon scales, the conformal Hubble rate $\mathcal H = a H$ determines
the characteristic time-scale associated with the evolution of the graviton field, so that $h^\prime \sim \mathcal H \, h$.
On the other hand, the medium response is dominated
by hard modes $k \sim T_{\rm rh}$ that live well inside the horizon.
For such modes, the combination
\begin{equation}
\delta = \frac{\omega^\prime}{\omega^2} = \frac{- \kappa h_{ij}^\prime k_i k_j}{2 \, \omega^3}
\sim \frac{\kappa h_{ij}^\prime}{k}
\sim \kappa h_{ij} \frac{\mathcal H}{T_{\rm rh}}
\label{WKBparameter}
\end{equation}
is a small parameter. Therefore, we may use a Wentzel–Kramers–Brillouin
(WKB) approximation (see e.g.~\cite{Bender:1999box} for a texbook treatment) 
to solve eq.~\eqref{eq:chi eom}.
This gives
\begin{equation}
\chi (\tau, \vec{k}) =
\left( \frac{\omega_{\rm rh}}{\omega} \right)^{\frac12}
\left[
	\chi_{\rm rh} \cos(\alpha)
	+
	\left( \frac{\chi_{\rm rh}^\prime}{\omega_{\rm rh}} 
		+ \frac{\omega_{\rm rh}^\prime \, \chi_{\rm rh}}{2 \omega_{\rm rh}^2} \right)
	\sin(\alpha)
	\right]
+ \mathcal O(\delta^2) \ ,
\label{ChiSolution}
\end{equation}
where
\begin{equation}
\alpha = \int_{\tau_{\rm rh}}^\tau \! d \tau^\prime \,
	\Omega(\tau')
	\, ,
\qquad\quad \text{and} \qquad\quad
\Omega = 
\omega + \frac{3\,\omega^{\prime \, 2}}{8\,\omega^3} - \frac{\omega^{\prime\prime}}{4\omega^2} + \mathcal O(\delta^3) \ .
\end{equation}
The functions $\chi_{\rm rh} (\vec{k}\,)$ and $\chi_{\rm rh}'(\vec{k}\,)$ are 
the scalar field and its conformal time derivative at the initial 
time $\tau_{\rm rh}$ (the end of reheating).
Likewise, $\omega_{\rm rh}$ and $\omega_{\rm rh}^\prime$ are the initial values 
of $\omega$ and $\omega^\prime$ given in~(\ref{eq:chi eom}).
The two-point function is therefore parametrized in terms of initial time 
correlators involving the field and its time derivatives;
see Eq.~(\ref{fullFT}) for the general expression.

In a universe that is approximately spatially homogeneous and isotropic, up to small corrections due 
to the super-horizon tensor modes,
one may parametrize these correlators as~\footnote{We discard the divergent 
vacuum contributions to these correlators; regardless of the
renormalization procedure, they will be suppressed by 
factors of $\kappa^2 H^2 / T^4$ relative to the thermal part, 
and therefore negligible.}
\begin{subequations}
\begin{align}
\big\langle \chi_{\rm rh} (\vec k \,) \chi_{\rm rh}{\!}^\dagger (\vec q \,) \big\rangle 
	&= 
	(2\pi)^3 \delta^3 (\vec q - \vec k \,) \frac{f_{\rm rh}}{\omega_{\rm rh}} 
	\ , 
\\
\big\langle \chi_{\rm rh}^\prime (\vec k \,) \chi_{\rm rh}{\!}^\dagger (\vec q \,) \big\rangle 
	&= 
	(2\pi)^3 \delta^3 (\vec q - \vec k \,)
		\frac{ f'_{\rm rh} }{2 \omega_{\rm rh}}
	\ , 
\\
\big\langle \chi_{\rm rh}^\prime (\vec k \,) \chi_{\rm rh}'{\!}^{\dagger} (\vec q \,) \big\rangle 
	&
	= (2\pi)^3 \delta^3 (\vec q - \vec k \,)
	\frac{f_{\rm rh}^{\prime\prime} + 4 \omega_{\rm rh}^2 f_{\rm rh}}{4 \, \omega_{\rm rh}} 
	\ ,
\end{align}
\label{ScalarInitialConditions}%
\end{subequations}
where the functions $f_{\rm rh}(\vec k) $, $f_{\rm rh}^\prime(\vec k)$, 
and $f_{\rm rh}^{\prime\prime} (\vec k)$ are the initial-time distribution 
function of the $\chi$ field and its conformal time derivatives.
If the plasma is initially in local thermal equilibrium, as 
is reasonable to assume according to the discussion in 
Sec.~\ref{subsec: Initial conditions},
the correlators \eqref{ScalarInitialConditions} are subject to a KMS 
symmetry~\cite{Bellac:2011kqa}. This implies the distribution function is 
equal to the Bose-Einstein distribution \eqref{BEdef} and depends 
only on the energy of on-shell modes.
At initial time, this energy is just equal to $\omega_{\rm rh}$,
so that the initial-time distribution function is given by
\begin{equation}
f_{\rm rh}(\vec k) = f_B(\omega_{\rm rh}) \ .
\label{fBlocal}
\end{equation}
Its derivatives are therefore suppressed by powers of the WKB 
parameter~(\ref{WKBparameter}),
\begin{align}
f_{\rm rh}^\prime 
	&= 
	\omega_{\rm rh}^\prime (\partial_\omega f)_{\rm rh} 
	\sim \delta 
	\ , 
&
f_{\rm rh}^{\prime\prime} 
	&= 
	\omega_{\rm rh}^{\prime\prime} (\partial_\omega f)_{\rm rh} 
	+
	\omega_{\rm rh}^{\prime \, 2} (\partial_\omega^2 f)_{\rm rh} \sim \delta^2 \ .
\label{frhDers}
\end{align}
Now we can compute the two contributions to the anisotropic stress 
in~\eqref{ScalarPiExplicit} and ~\eqref{ScalarPiImplicit}.
Using integral \eqref{I1}, one immediately obtains the explicit contribution
\begin{equation}
\pi_{ij}^{\rm expl.}
	=
	- \frac{\kappa^2}{3a^4} h_{ij}
	\int\! \frac{d^3\vec{k}}{(2\pi)^3} \, k f_B(k)
	=
	- \frac{\pi^2 \kappa^2 T^4}{90} h_{ij}
	=
	-
	2 H^2 h_{ij}
	\, .
\label{ScalarExplicit}
\end{equation}
After expanding to linear order in tensor perturbations,
the implicit contribution becomes
\begin{equation}
\pi_{ij}^{\rm impl.}
	=
	\frac{\kappa^2}{2a^4} \Pi_{ijkl}
	\int\! \frac{d^3\vec{k}}{(2\pi)^3} \frac{k_k k_l k_m k_n}{k^2}
\bigg(
\begin{multlined}[t]
\frac{f_B(k)}{ k } h_{mn} - \frac{\partial f_B(k)}{\partial k} h_{mn}^{\rm rh}
\\
- \frac{\sin[ 2k(\tau \!-\! \tau_{\rm rh})]}{2k^2} \frac{\partial [k f_B(k)] }{\partial k}
	h^{\prime \, \text{rh}}_{mn}
\bigg) \ .
\end{multlined}
\end{equation}
Using integrals~\eqref{I2} and \eqref{I4},
and neglecting the quickly decaying transient contribution
due to the sine term,
this gives the result
\begin{equation}
\pi_{ij}^{\rm impl.}
	=
	\frac{\pi^2\kappa^2T^4}{450}
	\big( h_{ij} + 4 h_{ij}^{\rm rh} \big)
	=
	\frac{2}{5} H^2
	\big( h_{ij} + 4 h_{ij}^{\rm rh} \big)
	\, .
\end{equation}
Adding the explicit contribution~\eqref{ScalarExplicit} to the above implicit contribution
one obtains the full anisotropic stress,
\begin{equation}
\pi_{ij} =  - \frac{2\pi^2 \kappa^2 T^4}{225} \big( h_{ij} - h_{ij}^{\rm rh} \big) 
	=
	- \frac{8}{5} H^2 \big( h_{ij} - h_{ij}^{\rm rh} \big) 
	\, .
\label{ScalarFinal}
\end{equation}
This result is the same as the kinetic theory results obtained in~\cite{Rebhan:1994zw,Weinberg:2003ur}.

\subsubsection*{Comparison with previous results}

If we assume that thermalization during reheating occurred on a globally
homogeneous and isotropic background space, the initial state would inherit
those symmetries, and would not depend on the primordial tensor modes
produced during inflation. Thus, the distribution function and its
derivatives appearing in~(\ref{ScalarInitialConditions}) would be
\begin{equation}\label{GlobalEquilibrium}
f_{\rm rh} (\vec k) = f_B(k) \ ,
\qquad\quad
f^\prime_{\rm rh} (\vec k) = f^{\prime\prime}_{\rm rh} (\vec k) = 0 \ .
\end{equation}
Assuming these initial conditions, instead of those in~\eqref{fBlocal} 
and~(\ref{frhDers}) corresponding to the local equilibrium state,
one obtains a different two-point function~\eqref{fullFT}. Consequently,
one also obtains a different implicit contribution to the anisotropic stress,
\begin{equation}
\pi_{ij}^{\rm impl.}
	=
	\frac{\kappa}{a^4} \Pi_{ijkl}
	\int\! \frac{d^3\vec{k}}{(2\pi)^3} \frac{k_k k_l}{\omega}
	\bigg[
	\frac{\omega_{\rm rh}}{k} \cos^2(\alpha)
	+
	\frac{k}{\omega_{\rm rh}}
	\sin^2(\alpha)
	\bigg]
	f_B(k)
	\, .
\label{ScalarGlobalImplicit}
\end{equation}
that, when expanded to linear order in tensor perturbation, reads
\begin{equation}
\pi_{ij}^{\rm impl.}
	=
	\frac{\kappa^2}{2a^4} \Pi_{ijkl}
	\int\! \frac{d^3\vec{k}}{(2\pi)^3} \frac{ k_k k_l k_m k_n }{ k^3 }
	f_B(k) \Big( h_{mn} - \cos[2k(\tau \!-\! \tau_{\rm rh})] h_{mn}^{\rm rh} \Big)
	\, .
\end{equation}
Using relations~\eqref{I2} and~\eqref{I5} to evaluate these integrals,
neglecting quickly decaying transients, and adding the result to the explicit contribution~\eqref{ScalarExplicit},
one thus obtains the complete anisotropic stress
\begin{equation}
\pi_{ij} =  - \frac{2\pi^2 \kappa^2 T^4}{225} h_{ij}
	=
	- \frac{8}{5} H^2 h_{ij}
	\, .
\label{ScalarGlobalPi}
\end{equation}
As expected, this expression is the same as \eqref{ScalarFinal}, except 
for the dependence on the initial tensor perturbation amplitude dropping
out.

The result in~(\ref{ScalarGlobalPi}) is at odds with the result obtained in Refs.~\cite{Ota:2025rll,Ota:2026yzp}. While these works report the correct 
anisotropic stress~\eqref{ScalarFinal}, which is consistent with the kinetic 
theory treatment, they argue that the initial state of the early universe plasma 
has to be globally homogeneous and isotropic, corresponding to the 
initial distribution being given by~(\ref{GlobalEquilibrium}).
Our analysis clearly shows such that such an initial state yields the 
anisotropic 
stress given in~\eqref{ScalarGlobalPi}. It then acts as an effective mass for the
tensor perturbation, and accordingly leads to a huge suppression of the
tensor perturbation amplitude. On the other hand, the local equilibrium 
state \eqref{ScalarInitialConditions}, which accounts for the presence of 
long-wavelength primordial tensor perturbations,
reproduces the expected near-conservation of super-horizon tensor modes.
We comment more on these disagreements in Sec.~\ref{sec:conclusion}.

Finally, we should comment on the original results from~\cite{Ota:2023iyh},
that reported an amplification of superhorizon tensor perturbations
interacting with a free-streaming scalar plasma. According to our analysis, 
this is not consistent with either global or local thermal equilibrium initial 
conditions. Rather, we attribute this discrepancy to the
computation in Sec.~3 of~\cite{Ota:2023iyh} that included some contributions 
that do appear in the hydrodynamic regime, but are absent in the free-streaming 
limit being considered in that section.

\subsection{Photons}
\label{subsec:photons}

The action of a free photon field in curved spacetime is given as
\begin{equation}
S[A_\mu,g_{\mu\nu}] = \int \! d^4 x \sqrt{-g} \left( 
	- \frac{1}{4} g^{\mu\rho} g^{\nu\sigma} F_{\mu\nu} F_{\rho\sigma} \right)
	\, ,
\end{equation}
where~$F_{\mu\nu} = \partial_\mu A_\nu - \partial_\nu A_\mu$ is the
field strength tensor.
The photon energy-momentum tensor
\begin{equation}
T^\mu{}_\nu = \Bigl( g^{\mu\rho} \delta_\nu^\sigma
	- \frac{1}{4} \delta^\mu_\nu g^{\rho\sigma} \Bigr) g^{\alpha\beta}
	F_{\rho\alpha} F_{\sigma\beta}
	\, ,
\label{photonEMT}
\end{equation}
sources the Einstein equation, and consequently contributes to the anisotropic 
stress in the equation of motion~\eqref{eq:graviton eom}
for tensor perturbations.

The explicit metric dependence in~\eqref{photonEMT} produces the explicit
contribution to the anisotropic stress~\eqref{ExplicitPi},
\begin{equation}
\pi_{ij}^{\rm expl.}
	=
    -
	\frac{\kappa^2}{a^4}
	\Pi_{ijkl} 
	\Big(
    \eta^{\mu\nu}
	\big\langle F_{\mu m} F_{\nu l} \bigr\rangle_{(0)} \delta_{kn}
	+
	\big\langle F_{k m} F_{l n} \big\rangle_{(0)}
	\Big)
	h_{mn}
    \, ,
\label{VectorExplicit}
\end{equation}
where the field strength tensor correlators in this expression need to be evaluated
to zeroth order in tensor perturbations.
The implicit contribution to the anisotropic stress derives from the linear perturbation
of the field strength tensor correlator,
\begin{equation}
\pi_{ij}^{\rm impl.}
	=
	\frac{\kappa}{a^4} \Pi_{ijkl} 
	\eta^{\alpha\beta}
    \big\langle F_{\alpha k} F_{\beta l} \big\rangle_{(1)}
	\, .
\label{VectorImplicit}
\end{equation}
To compute these correlators, one has to solve for the dynamics of the photon field,
which is determined by the perturbed Maxwell equation
\begin{equation} \label{PerturbedMaxwell}
\partial^\nu F_{\nu\mu} - \kappa \, h_{ij} \, \partial_i F_{j\mu}
=0
\, .
\end{equation}
Guided by the scalar example, we work to leading order in the super-horizon/adiabatic
expansion and neglect time derivatives of the tensor perturbation~$h_{ij}$ inside
the anisotropic stress. Keeping the first time derivative, as in the scalar case,
only affects subleading terms and does not change the final result.

The first step is to determine the two-point function of the field-strength tensor. 
We rewrite the perturbed Maxwell equation~\eqref{PerturbedMaxwell} in a second-order
form by acting with a derivative and using the Bianchi identity,\footnote{If one
wishes to work with the vector potential~$A_\mu$ directly, one must fix a gauge.
The Coulomb gauge used in~\cite{Ota:2026yzp}, or the conformal multiplier 
gauge~\cite{Huguet:2013dia,Glavan:2025iuw} are convenient choices, since they 
maintain conformal invariance and do not 
introduce explicit dependence on the scale factor.}
\begin{equation}
\big( \partial^2 - \kappa h_{ij} \partial_i \partial_j \big) F_{\mu\nu} = 0
	\, .
\label{BinachiMaxwellEq}
\end{equation}
Thus, to the relevant order, each component of~$F_{\mu\nu}$ satisfies the
same equation of motion as~\eqref{ChiEom} as does the conformally rescaled scalar field 
of the preceding example. It follows that the
field-strength correlator must be expressible as a derivative tensor structure
acting on a scalar two-point function. The original first-order Maxwell 
equation~\eqref{PerturbedMaxwell} further constrains this structure, fixing the correlator to
\begin{equation}
\big\langle F_{\mu\nu}(x) F_{\rho\sigma}(y) \big\rangle
	=
	4\partial^x_{[\mu} \big( \eta_{\nu][\sigma} + \kappa h_{\nu][\sigma} \big) \partial^y_{\rho]} 
	G(x;y)
    \, ,
\label{Vector2ptSolution}
\end{equation}
where brackets on indices stand for weighted anti-symmetrization. 
The scalar two-point function~$G(x;y)$ in~\eqref{Vector2ptSolution} 
obeys the same equation (in each argument) as the scalar-field correlator,
\begin{equation}
\big( \partial_x^2 - \kappa h_{ij} \partial^x_i \partial^x_j \big) G(x;y) = 0 \, ,
\end{equation}
and is therefore given by
\begin{equation}
G(x;y) = \big\langle \chi(x) \chi(y) \big\rangle
    = \int\! \frac{d^3\vec{k}}{(2\pi)^3} \frac{d^3\vec{q}}{(2\pi)^3} \, 
    e^{i \vec{k} \cdot \vec{x} - i \vec{q} \cdot \vec{y}} \,
    \big\langle \chi(\tau_x,\vec{k} \,) \chi^\dag(\tau_y, \vec{q}\,) \big\rangle
    \, ,
\label{Scalar2ptFT}
\end{equation}
where the Fourrier transformed correlator inside the integral is given by eq.~\eqref{fullFT}.
As in the scalar case, the correlator depends on the initial state of the plasma.
For the local equilibrium state defined by Eq.~\eqref{ScalarInitialConditions},
and neglecting time derivatives of the tensor perturbation, one obtains
\begin{align}
G(x;y)
    ={}& \int\! \frac{d^3\vec{k}}{(2\pi)^3} \, 
    e^{i \vec{k} \cdot (\vec{x} - \vec{y}) } \,
    \frac{ f_{\rm rh}(\vec{k}) }{ \sqrt{ \omega(\tau_x,\vec{k}\,) \, \omega(\tau_y,\vec{k}\,) } }
	\cos \big[ \alpha(\tau_x,\vec{k}\,) - \alpha(\tau_y,\vec{k}\,) \big]
    \, .
\label{Gxy}
\end{align}

Inserting the solution for the field-strength correlator~\eqref{Vector2ptSolution}
into the explicit contribution to the anisotropic stress~\eqref{VectorExplicit} yields a vanishing result,%
\footnote{The apparent
difference between this explicit contribution and the one reported in~\cite{Frob:2025sfq}
is solely due to different conventions and does not affect the final equation of motion for the tensor perturbations.
Here, we perturb the Einstein equation with one index raised, 
while~\cite{Frob:2025sfq} perturbed the Einstein equation with two lowered indices.}
\begin{equation}
\pi_{ij}^{\rm expl.}
	=
	0
	\, .
\end{equation}
Applying the same procedure to the implicit contribution~(\ref{VectorImplicit}),
and using the integrals~\eqref{I1}--\eqref{I3}, gives
\begin{align}
\pi_{ij}^{\rm impl.}
    ={}&
    \frac{4\kappa}{a^4} \Pi_{ijkl} 
    \Big[
	\eta^{\mu\nu} 
    \partial^x_{[\mu} \big( \eta_{k][l} + \kappa h_{k][l} \big) \partial^y_{\nu]} 
	G(x;y)
    \Big]_{(1)}
\nonumber \\
    ={}&
    \frac{\kappa^2 \pi^2 T^4}{225} h_{ij}
    -
    \frac{\kappa^2\pi^2 T^4}{45} h_{ij}
    +
    \frac{4 \pi^2 \kappa^2 T^4}{225} h_{ij}^{\rm rh}
    \, .
\label{PhotonImplThree}
\end{align}
We have written the final expression as a sum of three terms to highlight their
distinct origins. The first term arises from the $h_{ij}$-dependence of the scalar
two-point function~(\ref{Gxy}); the second term originates from the $h_{ij}$-dependence
of the tensor structure in the field-strength correlator~(\ref{Vector2ptSolution});
and the third term is inherited from the $h_{ij}^{\rm rh}$-dependence of the scalar
two-point function~(\ref{Gxy}). In other words, the first two contributions are
dynamical, while the third encodes the initial conditions. Combining all three
yields the total anisotropic stress
\begin{equation}
\pi_{ij} =
	- \frac{8}{5} H^2 \big( h_{ij} - h_{ij}^{\rm rh} \big) 
	\, .
\label{PhotonAnisotropic}
\end{equation}
which matches the result~(\ref{ScalarFinal}) obtained for the scalar plasma model.

\subsubsection*{Comparison with previous results}

Previous references~\cite{Frob:2025sfq,Ota:2026yzp} that considered a free-
streaming photon plasma assume that it is in a global thermal equilibrium state 
at the onset of the radiation era. Such a state does not contain
corrections due to the presence of tensor perturbations during reheating.
In our notation, this corresponds to using the initial state characterized by Eq.~\eqref{GlobalEquilibrium}
to evaluate the scalar propagator in Eq.~\eqref{Vector2ptSolution}.
As in the scalar case, this assumption eliminates
the dependence on the initial condition from~\eqref{PhotonImplThree}.
Thus we obtain the anisotropic stress
\begin{equation}
\pi_{ij} = - \frac{8}{5} H^2 h_{ij}\, ,
\label{PhotonGlobalAnisotropicStress}
\end{equation}
which takes the form of an effective graviton mass term, the same 
as~(\ref{ScalarGlobalPi}) obtained in the scalar case. This is the result that 
should have been obtained in Ref.~\cite{Ota:2026yzp} under the assumption of a 
global thermal state, rather than the Weinberg limit~\cite{Weinberg:2003ur}, 
which instead follows from assuming a local thermal state.

A natural question is why Ref.~\cite{Frob:2025sfq} did not obtain
Eq.~(\ref{PhotonGlobalAnisotropicStress}), given that the assumed initial
conditions are the same. We trace the discrepancy to the treatment of the
Maxwell dynamics: Ref.~\cite{Frob:2025sfq} solves the second-order equation
(\ref{BinachiMaxwellEq}), but the homogeneous part of that solution
must be fixed by imposing the original first-order constraint
(\ref{PerturbedMaxwell}). This step was not implemented, which removes a
homogeneous contribution associated with the tensor-perturbation-dependent
part of the tensor structure in~(\ref{Vector2ptSolution}). The missing term
contributes \(\Delta \pi_{ij}=2H^2 h_{ij}\) to the implicit anisotropic stress,
precisely accounting for the shift between the effective graviton mass
\(m_{\rm eff}^2=8H^2/5\) implied by~(\ref{PhotonGlobalAnisotropicStress}) and
the tachyonic value \(m_{\rm eff}^2=-2H^2/5\) reported in Ref.~\cite{Frob:2025sfq}.

\section{Discussion and conclusion}
\label{sec:conclusion}

The main goal of this work is to determine
whether super-horizon tensor perturbations are conserved during the radiation era, 
and how this depends on the state of the plasma produced by the
reheating period preceding it.
While hydrodynamics and older kinetic-theory results indicate that tensor perturbations freeze out,
several recent works have reported a deviation from super-horizon conservation in cases where
the primordial plasma contains free-streaming particles.

To settle this discrepancy, we explicitly computed the anisotropic stress
of the primordial plasma from first principles in two minimal models:
(i) a plasma of conformally coupled free scalars, and (ii) a plasma of free photons.
We argue that the physical state of early universe plasma after inflation and reheating is a {\it local} thermal equilibrium state.
This reflects the fact that local interactions thermalize the plasma on the background of long-wavelength tensor perturbations created during inflation.
This background modifies the dispersion relation of particles
in the early universe plasma, by making it dependent on the local
amplitude of super-horizon tensor perturbations at reheating $h_{ij}^{\rm rh}$. This 
prescription is contrary to the {\it global} thermal equilibrium state advocated
in~\cite{Ota:2025rll,Ota:2026yzp}, which supposed that thermalization happens on 
a homogeneous and isotropic background, in a process that is insensitive to the 
existence of primordial tensor perturbations.

In both models, local thermal
initial conditions yield the equation of motion for super-horizon 
tensor perturbations,
\begin{equation}
h''_{ij} + 2\mathcal{H}h'_{ij} 
	+ \frac{8}{5} \mathcal{H}^2 \bigl( h_{ij} - h_{ij}^{\rm rh} \bigr) = 0
	\, ,
\label{eq:local state equation}
\end{equation}
consistent with the prior kinetic theory computations \cite{Weinberg:2003ur,Baym:2017xvh}.
Note that, as discussed in Sec.~\ref{subsec: Initial conditions}, the anisotropic stress vanishes at~$\tau = \tau_{\rm rh}$
because the system is initially in local equilibrium, 
and becomes nonzero as it departs from equilibrium
after the onset of free-streaming.
Assuming radiation domination, one obtains for the solution 
of~(\ref{eq:local state equation})
\begin{equation}
\label{eq:local equilibrium solution}
h_{ij} = h_{ij}{\!\!}^{\rm rh}
	+
	e^{-N_{\rm rh}/2}
	\frac{h_{ij}'{\!}^{\rm rh}}{a_{\rm rh} H_{\rm rh}} 
	\frac{\sin( \lambda N_{\rm rh} )}{\lambda}
    \, ,
\end{equation}
where~$N_{\rm rh} = \ln(a/a_{\rm rh})$ is the number of e-foldings  since the 
onset of the radiation era, and~$\lambda = \sqrt{27/20}$.
The above solution clearly shows
that tensor perturbations 
are conserved on super-horizon scales,
as the decaying term is modulated by the time derivative of the
tensor amplitude at the end of reheating, which itself is already tiny.

In contrast,
a global equilibrium state is not initially impacted by the 
tensor perturbations, which in practice
amounts to taking $h_{ij}^{\rm rh} \to 0$ in the correlation functions that characterize the initial state of the plasma.
Hence, for both models, one obtains the equation of motion
\begin{equation}
h''_{ij} + 2\mathcal{H}h'_{ij} + \frac{8}{5} \mathcal{H}^2 h_{ij} = 0
	\, ,
\label{eq:global state equation}
\end{equation}
with an effective tensor mode mass~$m_{\rm eff}^2=8\mathcal{H}^2/5$,
which yields the solution
\begin{equation}
h_{ij} = e^{-N_{\rm rh}/2}
    \bigg[
    h_{ij}^{\rm rh} \cos(\lambda N_{\rm rh})
    +
    \Big( \frac{h_{ij}^{\rm rh}}{2}
        + \frac{ h_{ij}'{\!}^{\rm rh}}{a_{\rm rh} H_{\rm rh}} \Big)
        \frac{ \sin(\lambda N_{\rm rh})}{\lambda} 
    \bigg]
    \, .
\label{GlobalSuppression}
\end{equation}
This solution clearly shows a strong suppression of the tensor perturbation amplitude.
In other words, the evolution of super-horizon tensor perturbations
is extremely sensitive to the assumed initial state of the plasma.

To summarize, among the two prescriptions, it is only the local equilibrium
prescription that reproduces the standard super-horizon conservation
that underlies the standard mapping from inflation to observations at the CMB.
In contrast, the global prescription predicts a strong suppression the overall amplitude.%
\footnote{Early first-principles analyses of free-streaming 
scalars~\cite{Ota:2023iyh} and photons~\cite{Frob:2025sfq} reported a 
super-horizon growth of tensor modes; however, this growth does not survive a 
consistent treatment of the underlying dynamics and initial conditions. We trace 
the discrepancy to different issues in the two setups, as discussed at the end 
of Secs.~\ref{subsec:conformal scalars} and \ref{subsec:photons}.}

Given this result, it is natural to ask how Refs.~\cite{Ota:2025rll,Ota:2026yzp} arrive at Eq.~\eqref{eq:local state equation}, which is consistent with 
the kinetic-theory prediction~\cite{Weinberg:2003ur},
while simultaneously imposing a globally thermalized initial state.
Our reading is that the computation in those works first yields Eq.~\eqref{eq:global state equation}.
The agreement with the kinetic-theory result 
is then obtained by 
appealing to the argument
that super-horizon tensor modes should exhibit a shift symmetry,
and using this to shift
the tensor perturbation in Eq.~\eqref{eq:global state equation} so as to reproduce Eq.~\eqref{eq:local state equation}.
The coefficient of the shifted term is fixed
by additional symmetry arguments,
presented in Ref.~\cite{Ota:2025rll},
chosen so that the resulting equation coincides with the desired form.

While this procedure 
does lead
to a final equation that matches the expected kinetic-theory limit,
it raises several conceptual issues.
First, the shift symmetry in question is
at best an approximate property of the free tensor
kinetic operator in the super-horizon limit.
As emphasized already by Weinberg~\cite{Weinberg:2003ur},
it is not an actual gauge symmetry for modes of finite momentum.
This distinction matters because the
observationally relevant tensor modes 
necessarily have finite comoving momenta:
They are either physical perturbations with
well-defined amplitudes outside the horizon, or they are gauge artifacts.
Treating the super-horizon amplitude as pure gauge
would render it arbitrary and would
undermine the standard link between late-time tensor observables
and their inflationary origin.

A second set of concerns pertains to the treatment of initial conditions. Ref.~\cite{Ota:2025rll} argues that Ward identities
forbid imposing a locally thermalized state as an initial condition,
while simultaneously stating that local equilibrium
may nonetheless develop dynamically.
For a well-posed hyperbolic evolution problem, however,
the specification of initial data is part of the dynamics.
If a configuration is compatible with the equations and can arise
through evolution, it should also be admissible as initial data.
In the present context, we have shown explicitly that
a locally thermalized initial state is both physically
natural and consistent, and that it reproduces the
kinetic-theory result when propagated forward.
\\

Finally, it is worth emphasizing the scope of the present analysis.
Our computations focus on the super-horizon regime in radiation domination
and on free-streaming plasma species,
where hydrodynamics is not applicable.
Within this setting, the conclusion seems robust:
Once the initial plasma state is specified
locally on the perturbed background,
the induced anisotropic stress for gravitational waves does
not lead to an appreciable super-horizon evolution
beyond the expected decaying mode, and the standard
conservation of tensor perturbations is preserved.
We have not considered scalar cosmological perturbations,
which are of greater observational interest,
and it would be interesting to check whether
the same conclusions apply there as well.
It would also be 
interesting to examine 
far-from-equilibrium
situations that may occur during reheating,
either by modifying the particle distribution,
or in the more extreme case in regimes where
a quasi-particle picture cannot be applied directly.
Even with particles that are initially close to equilibrium
(as with warm inflation), one could envision scenarios
where a particle species freezes out before the
tensor modes are generated, potentially leading to an initial distribution
that is not in local thermal equilibrium. 

\

\

\begin{acknowledgements}\noindent
We are grateful to Atsuhisa Ota, Hui-Yu Zhu, and Yuhang Zhu for a useful
discussion on the contents of this paper.
DG was supported by project 24-13079S of the Czech Science Foundation (GAČR).
PK was supported by the Swiss National Science Foundation 
(SNSF) under grant Nr. P500PT-217885.  IS acknowledges support of the European Structural and Investment Funds and the Czech Ministry of Education, Youth and Sports (project FORTE --- CZ.02.01.01/00/22\_008/0004632). 
\end{acknowledgements}


\appendix

\section{Additional formulae}

\paragraph{General scalar two-point function.}
The scalar field two-point function for the scalar field solution in~(\ref{ChiSolution}) reads
\begin{align}
\MoveEqLeft[3]
\big\langle \chi (\tau_x, \vec{k}\,) \, \chi^\dag (\tau_y, \vec{q}\,) \big\rangle
    =
    \bigg[ \frac{\omega_{\rm rh}(\vec{k}\,) \, \omega_{\rm rh}(\vec{q}\,)}
        {\omega(\tau_x,\vec{k}\,) \, \omega(\tau_y,\vec{q}\,) }
        \bigg]^{\frac{1}{2} }
    \Bigg\{
	\big\langle \chi_{\rm rh}(\vec{k}\,) \chi_{\rm rh}^\dag(\vec{q}\,) \big\rangle
	\cos \big[ \alpha(\tau_x,\vec{k}\,) \big]
	\cos \big[ \alpha(\tau_y,\vec{q}\,) \big]
\nonumber \\
&
	+
	\bigg[
    \frac{ \big\langle \chi_{\rm rh}' (\vec{k}\,) 
        \chi^\dag_{\rm rh}(\vec{q}\,) \big\rangle }
            {\omega_{\rm rh}(\vec{k}\,) } 
	+ \frac{\omega_{\rm rh}'(\vec{k}\,) 
        \big\langle \chi_{\rm rh}(\vec{k}) \chi^\dag_{\rm rh}(\vec{q}\,) \big\rangle }
            {2 \omega_{\rm rh}^2( \vec{k}\,)} \bigg]
	\sin\big[ \alpha(\tau_x,\vec{k}\,) \big]
	\cos\big[\alpha(\tau_y,\vec{q}\,)\big]
\nonumber \\
&
	+
    \bigg[
    \frac{\big\langle \chi_{\rm rh}(\vec{k}\,)
        \chi_{\rm rh}'^{\, \dag}(\vec{q}\,) \big\rangle }
            {\omega_{\rm rh}(\vec{q}\,) } 
		+ \frac{\omega_{\rm rh}'(\vec{q}\,) 
            \big\langle \chi_{\rm rh}(\vec{k}\,)
                \chi^\dag_{\rm rh}(\vec{q}\,) \bigr\rangle }
            {2 \omega_{\rm rh}^2(\vec{q}\,) }
        \bigg]
	\cos\big[\alpha(\tau_x,\vec{k}\,) \big]
    \sin\big[\alpha(\tau_y,\vec{q}\,) \big]
\nonumber \\
&
	+
    \bigg[
	\frac{ \big\langle \chi_{\rm rh}'(\vec{k} \,)
        \chi_{\rm rh}'^{\,\dag}(\vec{q}\,) \big\rangle}
            {\omega_{\rm rh}( \vec{k} \,)
                \, \omega_{\rm rh}(\vec{q}\,)}
	+
    \frac{\omega_{\rm rh}'(\vec{q}\,)
        \big\langle \chi_{\rm rh}'(\vec{k} \,)
            \chi^\dag_{\rm rh}(\vec{q}\,) \big\rangle} 
            {2 \, \omega_{\rm rh}^2(\vec{q}\,) \,
                \omega_{\rm rh}( \vec{k} \,) } 
	+
    \frac{\omega_{\rm rh}'(\vec{k} \,) 
        \big\langle \chi_{\rm rh}(\vec{k} \,) 
            \chi_{\rm rh}'^{\,\dag}(\vec{q}\,) \big\rangle }
                {2 \, \omega_{\rm rh}^2(\vec{k} \,)
                    \, \omega_{\rm rh}(\vec{q}\,)}
\nonumber \\
&   \hspace{1cm}
	+
    \frac{\omega_{\rm rh}'(\vec{k} \,) \,
        \omega_{\rm rh}'(\vec{q}\,)
            \big\langle \chi_{\rm rh}(\vec{k} \,) 
                \chi_{\rm rh}^\dag(\vec{q}\,) \big\rangle}
                    {4 \, \omega_{\rm rh}^2(\vec{k} \,) \,
                        \omega_{\rm rh}^2(\vec{q}\,)}
	\bigg]
	\sin\big[\alpha(\tau_x, \vec{k}\,) \big] 
    \sin\big[\alpha(\tau_y, \vec{q}\,) \big] 
    \Bigg\}
    \, .
\label{fullFT}
\end{align}
It is parametrized in terms of the two-point functions between the field and its time
derivative at the initial (reheating) time, namely,
\begin{equation}
\big\langle \chi_{\rm rh}(\vec{k} \,) 
                \chi^\dag_{\rm rh}(\vec{q}\,) \big\rangle \, ,
\qquad
\big\langle \chi_{\rm rh}(\vec{k} \,) 
                \chi'^{\,\dag}_{\rm rh}(\vec{q}\,) \big\rangle \, ,
\qquad
\big\langle \chi'_{\rm rh}(\vec{k} \,) 
                \chi'^{\,\dag}_{\rm rh}(\vec{q}\,) \big\rangle \, .
\end{equation}
As is shown in Sec.~\ref{sec: Two free-streaming plasma models}, different choices
for these correlators lead to very different predictions.

\paragraph{Momentum space integrals.}
The computation in Sec.~\ref{sec: Two free-streaming plasma models} involves 
evaluating a number of momentum space integrals that involve the Bose-Einstein 
distribution $f_B(k) \!=\! \flatfrac1{( e^{\beta k} \!-\! 1 )}$.
The time-independent integrals are
\begin{align}
I_1
&=
\int\! \frac{d^3\vec{k}}{(2\pi)^3} \, \frac{k_i k_j}{k} \, f_B(k)
	=
    \delta_{ij} \times p
	\, ,
\label{I1}
\\
I_2
&=
\int\! \frac{d^3\vec{k}}{(2\pi)^3} \frac{ k_i k_j k_k k_l }{ k^3 } f_B(k)
	=
	\bigl( \delta_{ij} \delta_{kl} + \delta_{ik} \delta_{jl}
		+ \delta_{il} \delta_{jk} \bigr)
	\times \frac{p}5
	\, ,
\label{I2}
\\
I_3
&=
\int\! \frac{d^3\vec{k}}{(2\pi)^3} \frac{k_i k_j k_k k_l}{ k^2 }
	\frac{\partial f_B(k)}{\partial k}
	=
	\bigl( \delta_{ij} \delta_{kl} + \delta_{ik} \delta_{jl}
		+ \delta_{il} \delta_{jk} \bigr)
	\times - \frac{4 \, p}5
	\, ,
\label{I3}
\end{align}
where 
\begin{align}
p &= \frac13 \int\! \frac{d^3\vec{k}}{(2\pi)^3} \, k f_B(k) = \frac{\pi^2}{90\beta^4}
\end{align}
is the pressure due to a single bosonic degree of freedom in the early universe plasma.
The time-dependent integrals and their late-time limits are
\begin{align}
I_4
&=
\int\! \frac{d^3\vec{k}}{(2\pi)^3} \frac{k_i k_j k_k k_l}{k^4}
	\frac{\partial [k f_B(k)] }{\partial k}
	\sin(2k\Delta\tau)
\nonumber \\
&=
	\big( \delta_{ij} \delta_{kl} + \delta_{ik} \delta_{jl} + \delta_{il} \delta_{jk} \big)
	\times
	3 \beta \, p
	\bigg[
	\frac1{x^3} -
	\frac{2x + \sinh (2x) + x\cosh (2x)}{ \sinh^4 (x) }
	\bigg]
\nonumber \\
&\xrightarrow{x \to \infty}
	\big( \delta_{ij} \delta_{kl} + \delta_{ik} \delta_{jl} + \delta_{il} \delta_{jk} \big)
	\times 3 \beta \, p
	\bigg[
	\frac1{x^3} +
	\mathcal{O} \big( x \, e^{- 2x} \big)
	\bigg]
    \, ,
\label{I4}
\\
I_5
&=
\int\! \frac{d^3\vec{k}}{(2\pi)^3} \frac{k_i k_j k_k k_l}{k^3}
	f_B(k) \cos(2k\Delta\tau)
\nonumber \\
&=
	\big( \delta_{ij} \delta_{kl} + \delta_{ik} \delta_{jl} + \delta_{il} \delta_{jk} \big)
	\times - 9 \, p
	\bigg[
	\frac1{x^4} -
	\frac{1 + 2 \cosh^2(x)}{3 \sinh^4(x)}
	\bigg]
\nonumber \\
&\xrightarrow{x \to \infty}
	\big( \delta_{ij} \delta_{kl} + \delta_{ik} \delta_{jl}
		+ \delta_{il} \delta_{jk} \big)
	\times - 9 \, p
	\bigg[
	\frac1{x^4} -
	\mathcal{O} \big( \, e^{- 2 x} \big)
	\bigg]
	\, ,
\label{I5}
\end{align}
where~$x = 2\pi \Delta \tau/\beta$ is conformal time multiplied by the 
characteristic energy-scale of particles in the early universe plasma.
Thus, the limit $x \to \infty$ corresponds to considering macroscopic 
time-scales.
%



\bibliography{references}

@article{Hui:2018cvg,
    author = "Hui, Howard and others",
    editor = "Angeli, George Z. and Dierickx, Philippe",
    title = "{BICEP Array: a multi-frequency degree-scale CMB polarimeter}",
    eprint = "1808.00568",
    archivePrefix = "arXiv",
    primaryClass = "astro-ph.IM",
    doi = "10.1117/12.2311725",
    journal = "Proc. SPIE Int. Soc. Opt. Eng.",
    volume = "10708",
    pages = "1070807",
    year = "2018"
}

@article{SimonsObservatory:2025avm,
    author = "Abril-Cabezas, I. and others",
    collaboration = "Simons Observatory",
    title = "{The Simons Observatory: forecasted constraints on primordial gravitational waves with the expanded array of Small Aperture Telescopes}",
    eprint = "2512.15833",
    archivePrefix = "arXiv",
    primaryClass = "astro-ph.CO",
    reportNumber = "FERMILAB-PUB-25-0947-PPD",
    month = "12",
    year = "2025"
}

@article{Belkner:2023duz,
    author = "Belkner, Sebastian and Carron, Julien and Legrand, Louis and Umilt{\`a}, Caterina and Pryke, Clem and Bischoff, Colin",
    collaboration = "CMB-S4",
    title = "{CMB-S4: Iterative Internal Delensing and r Constraints}",
    eprint = "2310.06729",
    archivePrefix = "arXiv",
    primaryClass = "astro-ph.CO",
    doi = "10.3847/1538-4357/ad2351",
    journal = "Astrophys. J.",
    volume = "964",
    number = "2",
    pages = "148",
    year = "2024"
}

@book{Weinberg:2008zzc,
  title={Cosmology},
  author={Weinberg, Steven},
  isbn={9780191523601},
  year={2008},
  publisher={Oxford University Press}
}

@article{Amin:2014eta,
    author = "Amin, Mustafa A. and Hertzberg, Mark P. and Kaiser, David I. and Karouby, Johanna",
    title = "{Nonperturbative Dynamics Of Reheating After Inflation: A Review}",
    eprint = "1410.3808",
    archivePrefix = "arXiv",
    primaryClass = "hep-ph",
    doi = "10.1142/S0218271815300037",
    journal = "Int. J. Mod. Phys. D",
    volume = "24",
    pages = "1530003",
    year = "2014"
}

@book{Kolb:1990vq,
    author = "Kolb, Edward W. and Turner, Michael S.",
    title = "{The Early Universe}",
    reportNumber = "FERMILAB-BOOK-1990-01",
    doi = "10.1201/9780429492860",
    isbn = "978-0-429-49286-0, 978-0-201-62674-2",
    publisher = "Taylor and Francis",
    volume = "69",
    month = "5",
    year = "2019"
}

@book{Kapusta:2006pm,
    author = "Kapusta, J. I. and Gale, Charles",
    title = "{Finite-temperature field theory: Principles and applications}",
    doi = "10.1017/CBO9780511535130",
    isbn = "978-0-521-17322-3, 978-0-521-82082-0, 978-0-511-22280-1",
    publisher = "Cambridge University Press",
    series = "Cambridge Monographs on Mathematical Physics",
    year = "2011"
}

@article{Ghiglieri:2024ghm,
    author = {Ghiglieri, J. and Laine, M. and Sch{\"u}tte-Engel, J. and Speranza, E.},
    title = "{Double-graviton production from Standard Model plasma}",
    eprint = "2401.08766",
    archivePrefix = "arXiv",
    primaryClass = "hep-ph",
    reportNumber = "RIKEN-iTHEMS-Report-24, CERN-TH-2023-216",
    doi = "10.1088/1475-7516/2024/04/062",
    journal = "JCAP",
    volume = "04",
    pages = "062",
    year = "2024"
}

@article{Ringwald:2020ist,
    author = {Ringwald, Andreas and Sch{\"u}tte-Engel, Jan and Tamarit, Carlos},
    title = "{Gravitational Waves as a Big Bang Thermometer}",
    eprint = "2011.04731",
    archivePrefix = "arXiv",
    primaryClass = "hep-ph",
    reportNumber = "DESY 20-187, DESY-20-187, TUM-HEP-1293-20",
    doi = "10.1088/1475-7516/2021/03/054",
    journal = "JCAP",
    volume = "03",
    pages = "054",
    year = "2021"
}

@article{Ringwald:2022xif,
    author = "Ringwald, Andreas and Tamarit, Carlos",
    title = "{Revealing the cosmic history with gravitational waves}",
    eprint = "2203.00621",
    archivePrefix = "arXiv",
    primaryClass = "hep-ph",
    reportNumber = "DESY 22-037, TUM-HEP-1389-22",
    doi = "10.1103/PhysRevD.106.063027",
    journal = "Phys. Rev. D",
    volume = "106",
    number = "6",
    pages = "063027",
    year = "2022"
}

@article{Mukhanov:1981xt,
    author = "Mukhanov, Viatcheslav F. and Chibisov, G. V.",
    title = "{Quantum Fluctuations and a Nonsingular Universe}",
    journal = "JETP Lett.",
    volume = "33",
    pages = "532--535",
    year = "1981"
}

@article{Starobinsky:1979ty,
    author = "Starobinsky, Alexei A.",
    editor = "Khalatnikov, I. M. and Mineev, V. P.",
    title = "{Spectrum of relict gravitational radiation and the early state of the universe}",
    journal = "JETP Lett.",
    volume = "30",
    pages = "682--685",
    year = "1979"
}

@article{Parker:1968mv,
    author = "Parker, L.",
    title = "{Particle creation in expanding universes}",
    doi = "10.1103/PhysRevLett.21.562",
    journal = "Phys. Rev. Lett.",
    volume = "21",
    pages = "562--564",
    year = "1968"
}

@article{Planck:2018jri,
    author = "Akrami, Y. and others",
    collaboration = "Planck",
    title = "{Planck 2018 results. X. Constraints on inflation}",
    eprint = "1807.06211",
    archivePrefix = "arXiv",
    primaryClass = "astro-ph.CO",
    doi = "10.1051/0004-6361/201833887",
    journal = "Astron. Astrophys.",
    volume = "641",
    pages = "A10",
    year = "2020"
}

@article{LiteBIRD:2022cnt,
    author = "Allys, E. and others",
    collaboration = "LiteBIRD",
    title = "{Probing Cosmic Inflation with the LiteBIRD Cosmic Microwave Background Polarization Survey}",
    eprint = "2202.02773",
    archivePrefix = "arXiv",
    primaryClass = "astro-ph.IM",
    doi = "10.1093/ptep/ptac150",
    journal = "PTEP",
    volume = "2023",
    number = "4",
    pages = "042F01",
    year = "2023"
}

@article{Mukhanov:1990me,
    author = "Mukhanov, Viatcheslav F. and Feldman, H. A. and Brandenberger, Robert H.",
    title = "{Theory of cosmological perturbations. Part 1. Classical perturbations. Part 2. Quantum theory of perturbations. Part 3. Extensions}",
    reportNumber = "BROWN-HET-796, BROWN-HET-800, BROWN-HET-780",
    doi = "10.1016/0370-1573(92)90044-Z",
    journal = "Phys. Rept.",
    volume = "215",
    pages = "203--333",
    year = "1992"
}

@article{Weinberg:2003ur,
    author = "Weinberg, Steven",
    title = "{Damping of tensor modes in cosmology}",
    eprint = "astro-ph/0306304",
    archivePrefix = "arXiv",
    reportNumber = "UTTG-02-03",
    doi = "10.1103/PhysRevD.69.023503",
    journal = "Phys. Rev. D",
    volume = "69",
    pages = "023503",
    year = "2004"
}

@article{Ota:2023iyh,
    author = "Ota, Atsuhisa and Sasaki, Misao and Wang, Yi",
    title = "{One-loop thermal radiation exchange in gravitational wave power spectrum}",
    eprint = "2310.19071",
    archivePrefix = "arXiv",
    primaryClass = "astro-ph.CO",
    doi = "10.1007/JHEP03(2025)055",
    journal = "JHEP",
    volume = "03",
    pages = "055",
    year = "2025"
}

@article{Ota:2024idm,
    author = "Ota, Atsuhisa",
    title = "{Cosmological stimulated emission}",
    eprint = "2412.20474",
    archivePrefix = "arXiv",
    primaryClass = "astro-ph.CO",
    doi = "10.1140/epjc/s10052-025-14523-0",
    journal = "Eur. Phys. J. C",
    volume = "85",
    number = "7",
    pages = "813",
    year = "2025"
}

@article{Frob:2025sfq,
    author = {Fr{\"o}b, Markus B. and Glavan, Dra{\v{z}}en and Meda, Paolo and Sawicki, Ignacy},
    title = "{One-loop correction to primordial tensor modes during radiation era}",
    eprint = "2504.02609",
    archivePrefix = "arXiv",
    primaryClass = "astro-ph.CO",
    doi = "10.1007/JHEP12(2025)074",
    journal = "JHEP",
    volume = "12",
    pages = "074",
    year = "2025"
}

@article{Ota:2025yeu,
    author = "Ota, Atsuhisa and Zhu, Yuhang",
    title = "{Graviton stimulated emission in squeezed vacuum states}",
    eprint = "2504.06539",
    archivePrefix = "arXiv",
    primaryClass = "hep-th",
    doi = "10.1103/3mzq-lt9d",
    journal = "Phys. Rev. D",
    volume = "112",
    number = "10",
    pages = "103513",
    year = "2025"
}

@article{Ota:2026yzp,
    author = "Ota, Atsuhisa and Zhu, Hui-Yu and Zhu, Yuhang",
    title = "{Real-time Gravitational Wave Response in Thermal Spinning fields}",
    eprint = "2601.03631",
    archivePrefix = "arXiv",
    primaryClass = "hep-th",
    month = "1",
    year = "2026"
}

@article{Ota:2025rll,
    author = "Ota, Atsuhisa",
    title = "{Symmetry principles of gravitational perturbations in thermal environments}",
    eprint = "2510.22346",
    archivePrefix = "arXiv",
    primaryClass = "gr-qc",
    month = "10",
    year = "2025"
}

@article{Rebhan:1994zw,
    author = "Rebhan, Anton K. and Schwarz, Dominik J.",
    title = "{Kinetic versus thermal field theory approach to cosmological perturbations}",
    eprint = "gr-qc/9403032",
    archivePrefix = "arXiv",
    reportNumber = "DESY-94-040, TUW-93-23",
    doi = "10.1103/PhysRevD.50.2541",
    journal = "Phys. Rev. D",
    volume = "50",
    pages = "2541--2559",
    year = "1994"
}

@article{Liu:2024utl,
    author = "Liu, Leihua and Prokopec, Tomislav",
    title = "{Appearances are deceptive: can graviton have a mass?}",
    eprint = "2407.12657",
    archivePrefix = "arXiv",
    primaryClass = "hep-th",
    doi = "10.1007/JHEP05(2025)191",
    journal = "JHEP",
    volume = "05",
    pages = "191",
    year = "2025"
}

@article{Ghiglieri:2015nfa,
    author = "Ghiglieri, J. and Laine, M.",
    title = "{Gravitational wave background from Standard Model physics: Qualitative features}",
    eprint = "1504.02569",
    archivePrefix = "arXiv",
    primaryClass = "hep-ph",
    doi = "10.1088/1475-7516/2015/07/022",
    journal = "JCAP",
    volume = "07",
    pages = "022",
    year = "2015"
}

@article{Ghiglieri:2020mhm,
    author = "Ghiglieri, J. and Jackson, G. and Laine, M. and Zhu, Y.",
    title = "{Gravitational wave background from Standard Model physics: Complete leading order}",
    eprint = "2004.11392",
    archivePrefix = "arXiv",
    primaryClass = "hep-ph",
    doi = "10.1007/JHEP07(2020)092",
    journal = "JHEP",
    volume = "07",
    pages = "092",
    year = "2020"
}

@article{Drewes:2023oxg,
    author = "Drewes, Marco and Georis, Yannis and Klaric, Juraj and Klose, Philipp",
    title = "{Upper bound on thermal gravitational wave backgrounds from hidden sectors}",
    eprint = "2312.13855",
    archivePrefix = "arXiv",
    primaryClass = "hep-ph",
    doi = "10.1088/1475-7516/2024/06/073",
    journal = "JCAP",
    volume = "06",
    pages = "073",
    year = "2024"
}

@article{Frob:2025uev,
    author = {Fr{\"o}b, Markus B. and Glavan, Dra{\v{z}}en and Meda, Paolo},
    title = "{Measurements in stochastic gravity and thermal variance}",
    eprint = "2506.23193",
    archivePrefix = "arXiv",
    primaryClass = "gr-qc",
    month = "6",
    year = "2025"
}

@article{Hu:2008rga,
    author = "Hu, B. L. and Verdaguer, E.",
    title = "{Stochastic Gravity: Theory and Applications}",
    eprint = "0802.0658",
    archivePrefix = "arXiv",
    primaryClass = "gr-qc",
    doi = "10.12942/lrr-2008-3",
    journal = "Living Rev. Rel.",
    volume = "11",
    pages = "3",
    year = "2008"
}

@book{Hu:2020luk,
    author = "Hu, Bei-Lok and Verdaguer, Enric",
    title = "{Semiclassical and Stochastic Gravity}: {Quantum Field Effects on Curved Spacetime}",
    doi = "10.1017/9780511667497",
    isbn = "978-0-511-66749-7",
    publisher = "Cambridge University Press",
    address = "Cambridge",
    series = "Cambridge Monographs on Mathematical Physics",
    month = "1",
    year = "2020"
}

@book{Bender:1999box,
    author = "Bender, Carl M. and Orszag, Steven A.",
    title = "{Advanced Mathematical Methods for Scientists and Engineers I}",
    doi = "10.1007/978-1-4757-3069-2",
    publisher = "Springer",
    year = "1999"
}

@article{Baym:2017xvh,
    author = "Baym, Gordon and Patil, Subodh P. and Pethick, C. J.",
    title = "{Damping of gravitational waves by matter}",
    eprint = "1707.05192",
    archivePrefix = "arXiv",
    primaryClass = "gr-qc",
    reportNumber = "NORDITA-2017-069",
    doi = "10.1103/PhysRevD.96.084033",
    journal = "Phys. Rev. D",
    volume = "96",
    number = "8",
    pages = "084033",
    year = "2017"
}

@article{Huguet:2013dia,
    author = "Huguet, E. and Renaud, J.",
    title = "{Two-point function for the Maxwell field in flat Robertson-Walker spacetimes}",
    eprint = "1310.7333",
    archivePrefix = "arXiv",
    primaryClass = "hep-th",
    doi = "10.1103/PhysRevD.88.124018",
    journal = "Phys. Rev. D",
    volume = "88",
    number = "12",
    pages = "124018",
    year = "2013"
}

@article{Glavan:2025iuw,
    author = "Glavan, Dra{\v{z}}en",
    title = "{Two simple photon gauges in inflation}",
    eprint = "2503.12630",
    archivePrefix = "arXiv",
    primaryClass = "gr-qc",
    doi = "10.1007/JHEP06(2025)162",
    journal = "JHEP",
    volume = "06",
    pages = "162",
    year = "2025"
}

@book{Bellac:2011kqa,
    author = "Bellac, Michel Le",
    title = "{Thermal Field Theory}",
    doi = "10.1017/CBO9780511721700",
    isbn = "978-0-511-88506-8, 978-0-521-65477-7",
    publisher = "Cambridge University Press",
    series = "Cambridge Monographs on Mathematical Physics",
    month = "3",
    year = "2011"
}

@article{Lozanov:2019jxc,
    author = "Lozanov, Kaloian D.",
    title = "{Lectures on Reheating after Inflation}",
    eprint = "1907.04402",
    archivePrefix = "arXiv",
    primaryClass = "astro-ph.CO",
    month = "7",
    year = "2019"
}
\bibliographystyle{utphys}

\end{document}